\documentstyle[12pt]{article}
\begin{document}
\normalbaselineskip=16 true pt
\normalbaselines
\bibliographystyle{unsrt}
\setcounter{page}{0}

\def\be {\begin{equation}}
\def\ee {\end{equation}}
\def\bea {\begin{eqnarray}}
\def\eea {\end{eqnarray}}

\def\bbbar {B_d-\bar B_d}
\def\kkbar {K^0-\bar K^0}
\def\e {\epsilon}
\def\zbb {Z\rightarrow b\bar b}
\def\vll {{4\over 3}\eta_Bf_BB_Bm_B^2}
\def\thm {\tan\theta_H({\rm max})}

\thispagestyle{empty}
\rightline{\large\sf SINP-TNP/95-13}
\begin{center}
{\Large\bf
Constraints on a General Higgs Sector\\
from $\kkbar$, $\bbbar$ Mixing \\
and the $\e$ Parameter}
\end{center}
\vskip 1 true cm
\begin{center}
{\large\sf Debrupa Chakraverty} \footnote{E-mail: rupa@tnp.saha.ernet.in}
and {\large\sf Anirban Kundu} \footnote{E-mail: akundu@saha.ernet.in}\\[2mm]
Theory Group, Saha Institute of Nuclear Physics,\\
1/AF Bidhannagar, Calcutta - 700 064, India
\end{center}
\vskip 1 true cm

\begin{abstract}

The scalar sector of the Standard Model is extended to include
an arbitrary
assortment of scalars. In the case where this assignment does not preserve
$\rho=1$ at the tree-level, the departure from unity itself
puts the most stringent constraint
on the scalar sector, and where $\rho_{tree}=1$ is maintained, useful
bounds on the parameter space of the charged Higgs mass and
the doublet-nondoublet mixing angle can arise from data on $\bbbar$, $\kkbar$
mixing and the $\e$ parameter. These constraints turn out to be comparable
(and in some cases, better) to those obtained from $\zbb$ data.

\end{abstract}

\newpage

The electroweak symmetry breaking sector of the Standard Model (SM)
is still as cloudy as it was in the time of its formulation; and the main
factor responsible for this is the absence of any direct evidence of the
Higgs boson. The minimal version of the SM requires one complex scalar
doublet to break the electroweak symmetry; however, there is no a priori
reason why more scalars cannot exist. Models with two or more doublets have
been explored in this spirit \cite{hunter}.

It is also pertinent to investigate the consequences of scalars belonging
to non-doublet representations of $SU(2)$. This will enlarge the particle
content of the SM, and change the gauge-scalar as well as the fermion-scalar
interactions, without affecting the $SU(2)_L\times U(1)_Y$ gauge structure
of the model. That these non-doublet scalar representations can induce
Majorana masses for left-handed neutrinos has been shown \cite{gelmini}.
Collider signatures of scalars belonging to a triplet representation have
also been investigated \cite{collider}.

However, there is one serious constraint on these higher dimensional
($>2$) scalar representations: they in general do not maintain $\rho=1$
at tree-level. Singlet and doublet representations do not suffer from this
malady and that is why much work have been done on their phenomenological
implications \cite{grossman,gunion}.
For an arbitrary assortment of scalars, one
has three possibilities:

\begin{enumerate}

\item The higher dimensional multiplet does not {\em incidentally}
contribute to $\rho$. This will happen, e.g., for a multiplet with weak
isospin $T=3$ and weak hypercharge $Y=4$. However, being quite
artificial, such representations will not be discussed anymore in this
paper.

\item The vacuum expectation values (VEV) of the higher representations
are much smaller than the doublet VEV so that $\rho-1$ is within
experimental bound.

\item There is a remaining custodial $SU(2)$ symmetry among the higher
representations. In this case, the effects of the `bad' representations
on $\rho-1$ cancel out. For such a cancellation to remain valid even at
one-loop level, one requires a fine-tuning; however, it has been shown
\cite{gunion} that the fine-tuning required is of the same order as one
encounters in the SM. Following this prescription, some serious
model-building has been done in recent
times \cite{gunion, georgi}.

\end{enumerate}

Recently, a general formulation to treat arbitrary representations of
scalars was proposed \cite{km}. Only the constraint coming from the
tree-level absence of flavour-changing neutral currents (FCNC) was
assumed there. In simple terms, this constraint means that either a single
weak doublet $\Phi_1$ couples with both $T_3=+1/2$ and $T_3=-1/2$
fermions, or one weak doublet $\Phi_1$ couples with $T_3=+1/2$ and another,
$\Phi_2$, couples with $T_3=-1/2$ type fermions. For simplicity, we have
assumed that the same doublet couples with quarks and leptons.

It was shown in Ref. \cite{km} that if the arbitrary assortment of
multiplets do not keep $\rho=1$ at tree-level, then the constraint coming
from $\rho-1$ is by far the strictest to limit the doublet-nondoublet
mixing. (This mixing occurs because, in general, the weak and the mass basis
of scalars are not identical and states in these two basis are related by
some unitary matrices.) However, for those models which keep $\rho_{tree}
=1$ (either entirely consisting of doublets and singlets, or having
compensating `bad' representations --- possibility (3) as listed above),
a significant constraint on the parameter space of the singly-charged
Higgs mass $m_{H^+}$ and doublet-nondoublet mixing angle $\theta_H$
can be obtained from $\zbb$ data.

In this paper we investigate what constraints on the abovementioned
parameter space can be obtained from processes like $\bbbar$ and $\kkbar$
mixing, and from the experimental value of the $CP$-violating $\e$
parameter. Such a study was performed earlier for two-Higgs doublet
models \cite{thdm}.
We intend to show that the constraints sometimes turn out
to be better than those obtained with the $\zbb$ data.
As already pointed out by Grossman \cite{grossman}, other
processes do not play a significant role in constraining the parameter
space. We will show that for those models where the scalar sector contains
nondoublet representations, conclusions can differ significantly from the
ones drawn in the case of multi-Higgs doublet model. It may again be stressed
that we give a general treatment which yields the well-known results
for multi-Higgs doublet model at proper limit.

\bigskip\bigskip

Before proceeding further, let us set our notations, which will mostly
follow Ref. \cite{km}. In the weak basis, the Higgs multiplets are denoted
by $\Phi$ and the fields by $\phi$. In the mass basis, we use $H$ to denote
the fields. $H$ and $\phi$ are related via unitary matrices; for our purpose
it is sufficient to show a pair of such relations:
\bea
H_i^+&=&\alpha_{ij}\phi_j^+;\ \ \ \ \phi_i^+=
\alpha_{ji}^{\star}H_j^+;\nonumber\\
H_i^-&=&\alpha_{ij}^{\star}\phi_j^-;\ \ \ \ \phi_i^-=\alpha_{ji}H_j^-.
\eea
We also set $H_1^+\equiv G^+$, which means
\be
G^+=\sum_k\alpha_{1k}\phi_k^+.
\ee

Keeping the quarks in the weak basis, the Yukawa couplings are given by
\be
\bar u d H_i^+\ :\ {ig\over \sqrt{2}m_W}{\alpha_{i1}\over\alpha_{11}}
(m_uP_L-m_dP_R)
\ee
for the case where only $\Phi_1$ gives mass to both $u$- and $d$-type
quarks, and
\be
\bar u d H_i^+\ :\ {ig\over \sqrt{2}m_W}({\alpha_{i1}\over\alpha_{11}}
m_uP_L-{\alpha_{i2}\over \alpha_{12}}m_dP_R)
\ee
for the case where $\Phi_1$ ($\Phi_2$) gives masss to $u$($d$)-type
quarks. The projection operators are
\be
P_L=(1-\gamma_5)/2,\ \ \ \ P_R=(1+\gamma_5)/2.
\ee
These two models will henceforth be called Model 1 and Model 2
respectively. Consideration of quarks in the mass basis will introduce
the relevant elements of the quark mixing matrix.

\bigskip

In the SM, the short-distance part of $\Delta m_K$,
the $K_L-K_S$ mass difference, is given by
\be
\Delta m_K={G_F^2\over 6\pi^2}\eta_1m_KB_Kf_K^2|(V_{cd}^{\star}
V_{cs})|^2m_c^2I_1(x_c),
\ee
where $\eta_1$ takes care of the relevant short-distance QCD correction,
and $f_K$ is the kaon decay constant. $B_K$ parametrizes the error in using
vacuum insertion approximation to evaluate the matrix element
$<\bar K|\bar d\gamma^{\mu}(1-\gamma_5)s\bar d\gamma_{\mu}(1-\gamma_5)s|
K>$, and lies between 0 and 1. Using chiral perturbation theory as well as
hadronic sum rules, one obtains $B_K=1/3$ \cite{chipt}, whereas lattice QCD
studies give $B_K=0.85$ as the central value \cite{lattice}.
Other values like $B_K=0.70$
is obtained from $1/N$ expansion technique \cite{onebyn}.

The function $I_1(z)$ has the expression
\be
I_1(z)=1-{3z(1+z)\over 4(1-z)^2}-{3z^2\ln(z)\over 2(1-z)^3},
\ee
and for any quark $q$, we use $x_q=m_q^2/m_W^2$.

Parametrizing the quark mixing matrix in an approximate form \cite{chau}
\be
V_{CKM}=\pmatrix{1& s_{12}& s_{13}e^{-i\delta}\cr
 -s_{12}-s_{13}s_{23}e^{i\delta}& 1& s_{23}\cr
 s_{12}s_{23}-s_{13}e^{i\delta}& -s_{23}& 1},
\ee
where the cosines of the mixing angles have been approximated by unity
and $s_{13}$ is assumed to be one order of magnitude smaller than
$s_{12}$ and $s_{23}$, one obtains
\be
|(V_{cd}^{\star}V_{cs})|^2=s_{12}^2+s_{23}^4q^2
+2s_{12}s_{23}^2q\cos \delta
\ee
with $q=s_{13}/s_{23}$.

Expression for the $\bbbar$ mixing parameter, $x_d$, in the SM, is
\cite{kdd}
\be
x_d={\Delta m\over\Gamma}\Bigg\vert_{B_d}=\tau_b{G_F^2\over 6\pi^2}
\eta_BB_Bf_B^2m_Bm_t^2I_1(x_t)|V_{td}^{\star}V_{tb}|^2
\ee
where $\eta_B$ is the corresponding short-distance QCD correction.
$\sqrt{B_B}f_B$ is estimated from the lattice studies to be $0.14\pm
0.04$ GeV. $\tau_b|V_{td}^{\star}V_{tb}|^2$ can be written as
\be
\tau_b|V_{td}^{\star}V_{tb}|^2=\tau_b|V_{cb}|^2(s_{12}^2+q^2-2s_{12}
q\cos\delta ).
\ee

Lastly, the $CP$-violating parameter of the neutral kaon system, $\e$,
has the following expression in the SM:
\bea
|\e |&=&{G_F^2\over 12\pi^2}{m_K\over\sqrt{2}\Delta m_K} m_W^2B_K f_K^2
\Big[ \eta_1{\rm Im}(V_{cd}^{\star}V_{cs})^2x_cI_1(x_c)\nonumber\\
&{ }&+\eta_2{\rm Im}(V_{td}^{\star}V_{ts})^2x_tI_1(x_t)+2\eta_3
{\rm Im}(V_{cd}^{\star}V_{cs}V_{td}^{\star}V_{ts})x_cI_2(x_c,x_t)\Big],
\eea
where, apart from the symbols previously explained,
\be
I_2(z_1,z_2)=\ln (z_2/z_1)-{3\over 4}{z_2\over 1-z_2}\Big[ 1+
{z_2\over 1-z_2}\ln z_2\Big].
\ee
$\eta_1$, $\eta_2$ and $\eta_3$ are three QCD correction factors, $\eta_1$
being the same as in eq. (6).

\bigskip

Now let us concentrate on the contributions to the abovementioned
parameters coming from an extended Higgs sector. Our discussion will
be limited within those assortment of scalar multiplets which keep
$\rho_{tree}=1$; however, a generalization is straightforward but of
little physical importance.

One has to consider two new box diagrams: one with two charged Higgses
and two up-type quarks, and one with one charged Higgs, one $W^+$
and two up-type quarks. Note that as per eq. (2), the new physics
contribution should exclude the diagrams containing only $H_1^+$
and no other charged Higgses.

To avoid cumbersome formulae which do not shed much light to new physics
issues, we assume all charged Higgses to be degenerate in mass \cite{km}.
This is not a too drastic approximation if one considers the fact that
it is the mass of the charged Higgs, $m_{H^+}$, which we want to
constrain. In case the charged Higgses do not have the same mass,
$m_{H^+}$ corresponds to the lightest physical one. To do meaningful
numerology, one has either to assume that all $H^+$s are degenerate, or
that one of them is light enough to conribute and the others are so
heavy that they effectively decouple. However, physically interesting
models \cite{georgi} do have all scalar masses of the same order of
magnitude, and so we stick to the first approximation. It may be
mentioned that if the masses of the charged scalars are not exactly the
same but similar in magnitude, bounds that we obtain change very little.
We will also state what happens if one considers the second limit, i.e.,
existence of only one `light' charged scalar.

Another reasonable approximation is to take all other quarks except the top
to be massless while considering their couplings to the scalar fields.
This makes eqs. (3) and (4) identical, and the results thus obtained will
be more general. Note that as the scalar coupling to fermion-antifermion
pair is proportional to the fermion mass, the GIM mechanism is not
operative.

We give expressions for the contributions of the scalar-mediated diagrams
to $\Delta m_K$, $x_d$ and $\e$. For any general quark $q$, we use
$y_q=m_q^2/m_{H^+}^2$. As all $m_{H^+}$s are assumed to be same, $y_q$
is unique.

The contribution to $\Delta m_K$ is
\be
\Delta m_K^H={G_F^2\over 24\pi^2}\eta_1m_KB_Kf_K^2|V_{td}^{\star}V_{ts}|
^2m_t^2(J_{HH}+J_{HW})
\ee
where
\be
J_{HH}=\sum_{i,j}\prime~y_t{\alpha_{i1}^2\alpha_{j1}^2\over\alpha_{11}^4}
\Big[ {1+y_t\over (1-y_t)^2}+{2y_t\ln y_t\over (1-y_t)^3}\Big],
\ee
and
\be
J_{HW}= \sum_{i=2}^n ~x_t\Big({\alpha_{i1}^2\over\alpha_{11}^2}\Big)
\Big[ 2I_3(x_t,x_H)-8I_4(x_t,x_H)\Big].
\ee
Here $\sum '$ means that the sum over both the mass indices runs from
1 to $n$, the number of charged scalars, but $i=1$, $j=1$ term corresponding
to the Goldstone contribution is to be subtracted, as that is already
considered in the SM amplitude. The same logic applies for the sum in
eq. (16). The expression for the two functions, $I_3$ and $I_4$, are
given by
\bea
I_3(x_H,x_t)&=&{x_t\over (1-x_t)(x_H-x_t)}-{x_H^2\ln x_H\over (1-x_H)
(x_t-x_H)^2}\nonumber\\
&{ }&+{x_t(2x_H-x_t-x_tx_H)\ln x_t\over (x_H-x_t)^2(1-x_t)^2},\\
I_4(x_H,x_t)&=&-{1\over (1-x_t)(x_H-x_t)}+{x_H\ln x_H\over (1-x_H)
(x_t-x_H)^2}\nonumber\\
&{ }&-{(x_H-x_t^2)\ln x_t\over (x_H-x_t)^2(1-x_t)^2},
\eea
with $x_H=m_{H^+}^2/m_W^2$. Note that $I_3$ differs in sign
from that given in
eq. (B.3) of Ref. \cite{grossman}.

$x_d$ is enhanced by
\be
x_d^H=\tau_b{G_F^2\over 24\pi^2}\eta_BB_Bf_B^2m_Bm_t^2|V_{td}^{\star}
V_{tb}|^2(J_{HH}+J_{HW}),
\ee
and the contribution to $\e$ is
\be
|\e |^H={G_F^2\over 48\pi^2}{m_K\over\sqrt{2}\Delta m_K}m_W^2B_Kf_K^2
m_K\eta_2x_t{\rm Im}(V_{td}^{\star}V_{ts})^2(J_{HH}+J_{HW}).
\ee

None of the above charged scalar mediated processes are possible if
$\alpha_{i1}=0$ for $i\not= 1$. In other words, the charged scalar of
the weak doublet that gives mass to the top quark must mix with charged
scalars of other multiplets to produce such contributions. This mixing
is parametrised by $\theta_H$, i.e., $\alpha_{i1}=\cos\theta_H$. From
the unitarity of the $\alpha$ matrix, $\sum_{i=2}^n |\alpha_{i1}|^2
=\sin^2\theta_H$.

Thus, if all $H^+$s are degenerate, $J_{HH}$ is proportional to
$\sec^4\theta_H-1$ and $J_{HW}$ is proportional to $\sec^2\theta_H
-1$. However, if only the $k$-th charged scalar effectively contributes,
the element $|\alpha_{k1}|^2$, and not the sum, gets paramount
importance. It may happen that $|\alpha_{k1}|^2$ is very small or
actually zero. Such a thing happens if $H_5^+$ is the lightest charged
scalar in the triplet model of Ref. \cite{georgi}. In this case, all
our discussions are invalidated, and we arrive at the well-known result
of possible existence of a light charged scalar which does not couple to
fermions.

Assuming the degeneracy of $m_{H^+}$, we try to put constraints on
$m_{H^+}-\tan\theta_H$ plane. A major obstacle in that direction is
the fact that a lot of quantities like $B_K$, $B_Bf_B^2$, $s_{23}$, $\delta$,
and even $m_t$, are poorly known or estimated. To be consistent with
the present experimental data, we take \cite{pdg, cdf}
\bea
G_F&=&1.16639\times 10^{-5}~{\rm GeV}^{-2},\ \  \Delta m_K=3.52\times
10^{-15}~{\rm GeV},\nonumber\\
m_{B_d}&=&5.28~{\rm GeV},\ \ m_t=176~{\rm GeV},\ \ m_W=80.41
{}~{\rm GeV},\nonumber\\
x_d&=&0.77,\ \ |\e |=2.26\times 10^{-3},\ \ f_K=0.165~{\rm GeV},\ \
m_K=0.498~{\rm GeV},\nonumber\\
s_{12}&=&0.2205,\ \ s_{23}=0.040,\ \ \tau_b|V_{cb}|^2=3.5\times 10^{-9}
{}~{\rm GeV}^{-1}.
\eea
The numerical values of the QCD correction factors that we use are
\cite{datta, quinn}
\be
\eta_1=0.78,\ \ \eta_2=0.60,\ \ \eta_3=0.37,\ \ \eta_B=0.85.
\ee

First, let us concentrate on $\Delta m_K$. Assuming no long-distance
contribution, $\Delta m_K$ does not limit $\tan\theta_H$ significantly.
For $B_K=1/3$, the maximum value of $\tan\theta_H$ is $7.4$, $6.2$ and
$7.6$ for $m_{H^+}=100$, 200 and 500 GeV respectively. This bound is one
order of magnitude poorer than that derived from $\zbb$ data. Though
formally eqs. (6) and (10) contain $\delta$, the result is insensitive to
its specific value; the reason is the small coefficient of $\cos\delta$
in eq. (9). For $B_K=0.85$, the bounds are a shed better: $\thm = 3.5$,
$2.9$ and $3.6$ for $m_{H^+}=100$, 200 and 500 GeV. We note that the bound
is `strongest' at around $m_{H^+}=m_t$.

The situation is different if one has, say, a 50\% long-distance
contribution. In that case, $B_K=1/3$ gives $\thm = 4.5$, $3.7$ and $4.6$
for $m_{H^+}=100$, 200 and 500 GeV. However, $B_K=0.85$ oversaturates
the SM value and no room for new physics is left. $B_K=0.70$ yields a
fairly strong constraint: $\thm = 1.2$, $1.0$ and $1.2$ for the three
values of $m_{H^+}$ we have chosen to mention. This is comparable to
those bounds obtained from partial width of $Z$ into $b\bar b$ pairs.

With $B_K=1/3$, $s_{23}=0.040$ and $q=0.10$, the strongest bound on
$\thm$ is 1.1, which is for $\delta=7\pi/12$ and $m_{H^+}=300$ GeV.
For $q=0.06$, the bound is somewhat less stringent; the results are
shown in Figs. (1a) and (1b). Also, $q=0.14$ constrains the parameter
space more tightly. Furthermore, one observes that $B_K=0.85$
does not allow $\delta > \pi/4$, and $B_K=0.70$ does not allow
$\pi/4 < \delta < 3\pi/4$ --- the SM value saturates the experimental
number. Even for those values of $\delta$ which allows for a new physics
contribution, $\thm$ is generally less than 1.0, which is a better
constraint than that obtained from $\zbb$ data.

Currently favoured values of $B_Bf_B^2$ ($\approx 0.02$ GeV$^2$) also
does not allow $\delta > \pi/2$ from measurements on $x_d$. For $\delta
=\pi/2$, one gets $\thm = 0.36$ for $m_{H^+}=100$ GeV. Even for $\delta$
as low as $\pi/6$, $\thm = 1.4$ for a charged scalar mass of 100 GeV.
Lowering $B_Bf_B^2$ to $0.01$ GeV$^2$ results in  a larger allowed value
of $\thm$. Figs. (2a) and (2b) show the detailed result.

We conclude from this analysis that even in a model with arbitrary assortment
of scalars, one can obtain fairly strong constraints on the parameter
space of the scalar sector, with a very few reasonable assumptions,
from $\bbbar$ mixing data and the $\e$ parameter, and maybe even from
$\kkbar$ mixing data. These constraints are shown to be comparable, and
sometimes better, to those obtained from $\Gamma(\zbb )$, which was
calculated in Ref. \cite{km}. We want to remind our readers that such an
analysis is only meaningful if the lightest physical charged scalar(s)
couple with fermions, and if $\rho_{tree}=1$ is maintained (otherwise,
$\rho$ parameter puts a better constraint). The error bar in $m_t$ turns
out to be insignificant; however, quantities like $B_K$, $B_Bf_B^2$ and
$\delta$, which are either poorly known or completely unknown, play a
significant role. With a more accurate experimental determination of
these quantities, one hopes to make these constraints more
meaningful.

\bigskip\bigskip

The authors thank T. De, B. Dutta-Roy and B. Mukhopadhyaya for useful
discussions. AK thanks International Centre for Theoretical Physics,
Trieste, Italy, for its hospitality, where a large part of the work
was done.

\newpage

\centerline{\large\bf Figure Captions}

{\bf 1(a)}. Upper limits on tan$\theta_H$ for different values of $m_{H^+}$,
as obtained from the analysis of the $\e$ parameter. We take $q=0.10$.
The uppermost curve is for $\delta=\pi/6$, and the successive ones are for
$\delta=\pi/4$, $\pi/3$, $5\pi/12$, $3\pi/4$ and $7\pi/12$ respectively.

{\bf 1(b)}. Same as in 1(a), with $q=0.06$. The curves are for
$\delta=\pi/6$, $\pi/4$,
$\pi/3$, $3\pi/4$, $5\pi/12$ and $7\pi/12$ respectively.

{\bf 2(a)}. Upper limits on tan$\theta_H$ for different values of $m_{H^+}$,
as obtained from the analysis of $B_d-\bar B_d$ mixing. We take
$B_Bf_B^2=0.02$.
The uppermost curve is for $\delta=\pi/6$, and the successive ones are
for $\delta=\pi/4$, $\pi/3$, $5\pi/12$ and $\pi/2$ respectively.
For $\delta>\pi/2$, the SM value saturates the experimental bound.

{\bf 2(b)}. Same as in 2(a), with $B_Bf_B^2=0.01$. The curves are for
$\delta=\pi/6$, $\pi/4$, $\pi/3$, $5\pi/12$, $\pi/2$, $7\pi/12$,
$2\pi/3$, $3\pi/4$ and $5\pi/6$ respectively.

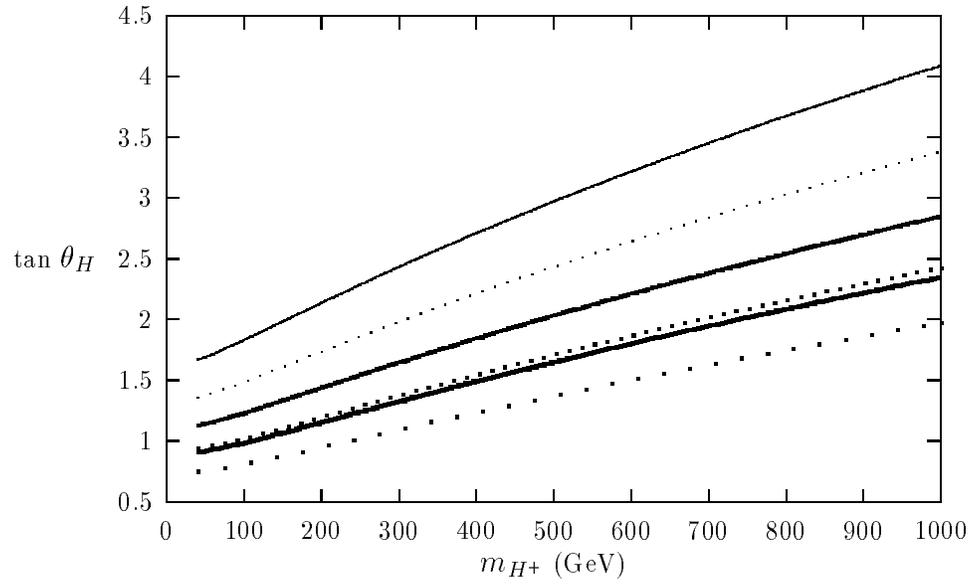
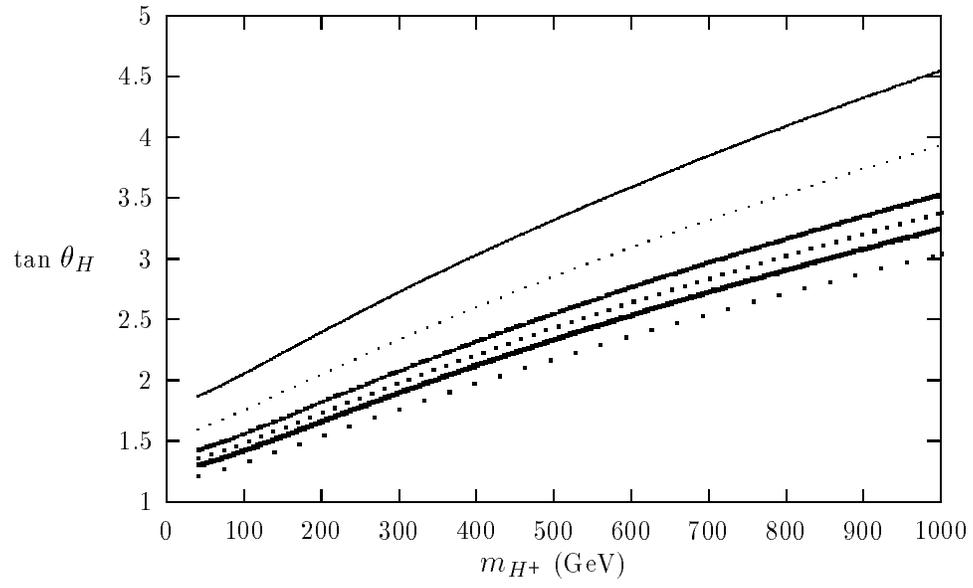
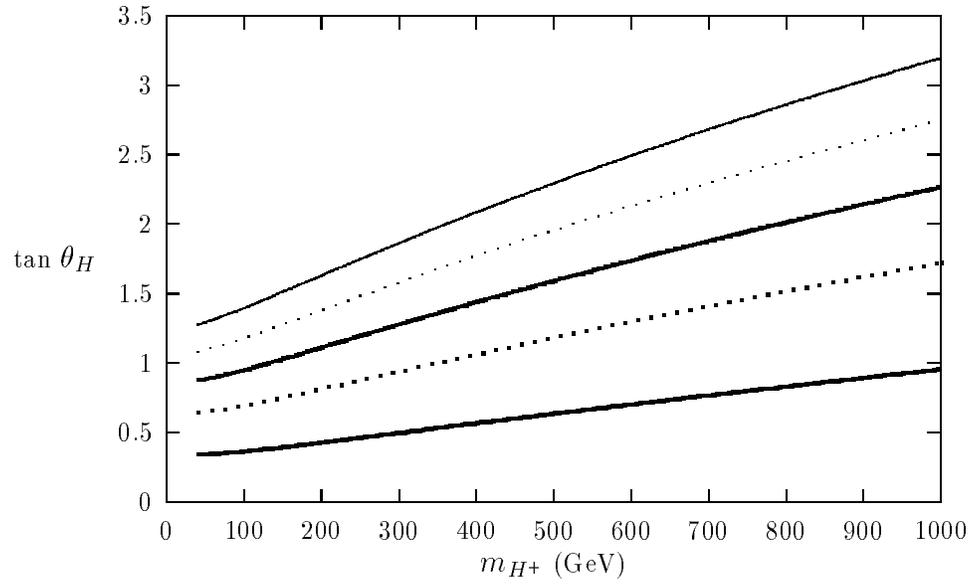
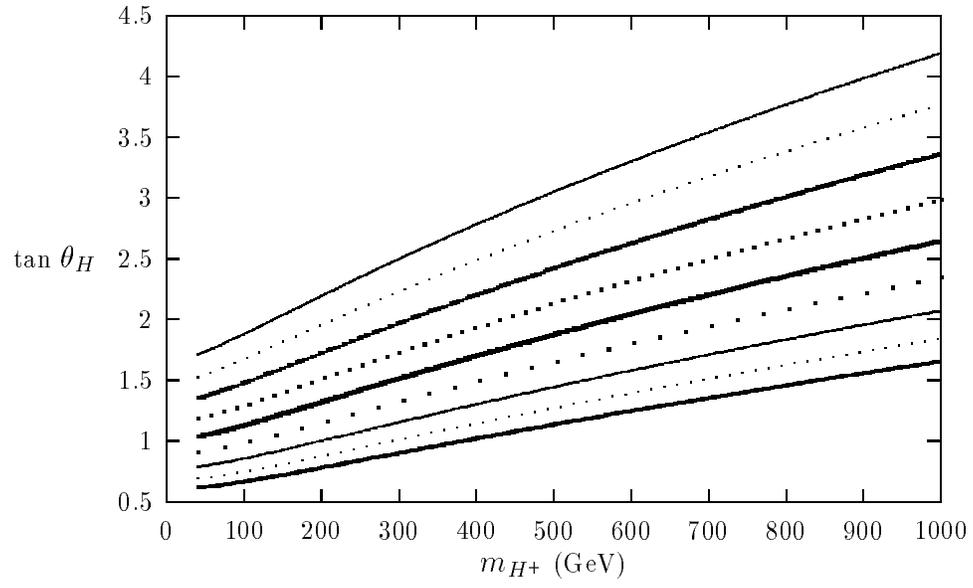
\end{document}